# Bayesian reconstruction of HIV transmission trees from viral sequences and uncertain infection times

Hesam Montazeri[1,2], Susan Little [3], Niko Beerenwinkel[1,2], Victor DeGruttola [4*],

1 Department of Biosystems Science and Engineering, ETH Zurich, Basel, Switzerland
2 SIB Swiss Institute of Bioinformatics, Basel, Switzerland
3 Department of Medicine, University of California San Diego, California, USA
4 Harvard School of Public Health, Boston, USA

* degrut@hsph.harvard.edu

## Abstract

Genetic sequence data of pathogens are increasingly used to investigate transmission dynamics in both endemic diseases and disease outbreaks; such research can aid in development of appropriate interventions and in design of studies to evaluate them. Several methods have been proposed to infer transmission chains from sequence data; however, existing methods do not generally reliably reconstruct transmission trees because genetic sequence data or inferred phylogenetic trees from such data are insufficient for accurate inference regarding transmission chains. In this paper, we demonstrate the lack of a one-to-one relationship between phylogenies and transmission trees, and also show that information regarding infection times together with genetic sequences permit accurate reconstruction of transmission trees. We propose a Bayesian inference method for this purpose and demonstrate that precision of inference regarding these transmission trees depends on precision of the estimated times of infection. We also illustrate the use of these methods to study features of epidemic dynamics, such as the relationship between characteristics of nodes and average number of outbound edges or inbound edges– signifying possible transmission events from and to nodes. We study the performance of the proposed method in simulation experiments and demonstrate its superiority in comparison to an alternative method. We apply them to a transmission cluster in San Diego and investigate the impact of biological, behavioral, and demographic factors.

## Introduction

Molecular epidemiology is increasingly used to investigate outbreaks or endemic diseases. Field studies of contacts between individuals that are capable of transmitting diseases also provide useful information in such settings, but can be difficult to collect when the nature of the contact required for transmission touches on sensitive issues, such as in the setting of sexually transmitted infections. This issue, along with the decreasing cost of genome sequencing, has led to the increasing use of molecular epidemiology in outbreak analysis Cottam et al. (2008); Didelot et al. (2014); Gilchrist et al. (2015); Lau et al. (2015); Volz et al. (2013). In particular, there is growing interest in developing computational models to identify transmission history or patterns for infectious disease outbreaks or for endemic diseases. It has been shown, for example, that knowledge of transmission history provides valuable information to guide public health interventions Ferguson et al. (2001); Keeling et al. (2003).

A variety of computational methods have been proposed to infer the history of transmissions among hosts using genetic data Cottam et al. (2008); Didelot et al. (2014); Lau et al. (2015); Mollentze et al. (2014); Morelli et al. (2012); Ypma et al. (2012, 2013). These methods can be divided into two main categories. The first is based on genetic distance Jombart et al. (2011); Smith et al. (2009); Snitkin et al. (2012); Spada et al. (2004); Wertheim et al. (2011,?). In these methods, small genetic distance between two hosts, e.g. less than 1% of sequence length, are believed to imply a transmission or at least membership in the same transmission subnetwork. The second is based on likelihood of parameters that characterize transmission trees for the genetic and other data available from an outbreak Jombart et al. (2014); Mollentze et al. (2014); Morelli et al. (2012); Ypma et al. (2012). Most of these methods simultaneously estimate phylogenetic and transmission trees Cottam et al. (2006, 2008); Didelot et al. (2014); Lau et al. (2015); Romero-Severson et al. (2014); Ypma et al. (2013).

The relationship between phylogenetic and transmission trees has been an active area of recent research Kenah et al. (2016); Romero-Severson et al. (2014); Ypma et al. (2013). It has been shown that there is no one-to-one match between phylogeny and transmission history Pybus and Rambaut (2009); Romero-Severson et al. (2014); Worby et al. (2014a,b). In particular, the topology of a phylogeny may be entirely different from the topology of the corresponding transmission tree Kenah et al. (2016); Leventhal et al. (2012); Worby et al. (2014b). It has also been demonstrated that the timing and order of transmission events are not generally inferable from a phylogenetic tree Romero-Severson et al. (2014). Kenah et al. showed that there are at most $2^{n-1}$ transmission trees consistent with a phylogenetic tree with $n$ leaves Kenah et al. (2016). However, incorporation of additional information collected during an outbreak such as locations or times of infections can substantially reduce the number of possible transmission trees.

In this paper, we propose a novel Bayesian method that incorporates genetic data and infection times known with error for inference of transmission trees. We demonstrate that knowledge of time of infection as well as genetic sequence data are necessary for an accurate inference of transmission trees. Although exact infection times are rarely known, intervals of infection can often be established from repeated testing or from HIV recency assays Janssen et al. (1998); Kothe et al. (2003); our methods were developed to accommodate such information. In simulation studies, we demonstrate the accuracy of the proposed method in reconstructing the true transmission increases as the length of the infection interval decreases. We also demonstrate that without such information or when infection is known only to within large time intervals, inference on underlying transmission trees is highly unreliable.

Information regarding infection intervals is often available or can be established from recency assays applied to stored samples. The proposed methods permit infection intervals to vary in width. In the HIV setting, patients experiencing primary infection—detectable at diagnosis—are known to have been infected within four months Moss and Bacchetti (1989). Recency assays permit inference about the intervals of infection up to a period of 2 years prior to the test. Further information about estimation intervals is available from analysis of genetic sequence Kouyos et al. (2011); Taffe and May (2008). These uncertain infection times in addition to observed sequences are the data required for implementation of our method.

The outline of this paper is as follows. Section 2 proposes a Bayesian analysis method for transmission tree inference. Section 3 presents results of a simulation study. Section 4 provide results of investigation of the performance of the inference method on an HIV dataset from San Diego. Section 5 provides conclusions.

## Materials and methods

In this section, we first study the relationship between phylogenetic and transmission trees and then propose a Bayesian inferential method for estimating transmission trees. Throughout the paper, we make the following assumptions:

1. Each patient is infected exactly once; superinfection is not modeled in the proposed approach.

2. Each infection begins by a single pathogen strain. After a certain period of within-host evolution, the evolved pathogen infects other patients, i.e. pathogen diversity within patients is not modeled.

3. All infected individuals are observed in the population and there is a pathogen sequence for each individual (no missing observations). In particular, each infected person (except the first infected) has an infector in the given observed population.

4. After sequencing, infected individuals do not infect other individuals due to change of behavior or effective treatments.

Similar assumptions have been made in other published studies. For example, the first three assumptions were made in Kenah et al. (2016) and the fourth, in Kühnert et al. (2016). As mentioned above, our proposed method requires knowledge regarding the intervals in which infections occur, e.g., a 95% confidence interval for an infection time. An infection interval that is arbitrarily wide implies that no information is available about the time of infection.

We show that making use of infection time information in the estimation process results in significant improvements in reconstructing a transmission tree. Our method accommodates lack of such information for some patients, although this situation will decrease performance.

We denote $I_p$ and $S_p$ as infection and sequencing times of patient $p \in \mathcal{P}$, respectively, where $\mathcal{P}$ denotes the set of observed infected individuals. We define the most recent sequencing time among all observed patients as the reference time point and set it to zero. We define all other time points backwards in calendar time with respect to this reference point. We assume a single pathogen sequence, denoted by $g_p$, is available for each $p \in \mathcal{P}$. A transmission tree $T$ is defined as the set of all transmissions between individuals in $\mathcal{P}$. A transmission $p \to q$ in $T$ indicates $p$ infects $q$. For example, the transmission tree $\{A \to B, A \to C\}$ means $A$ infected both $B$ and $C$. A phylogenetic tree, denoted by $P$, represents the evolutionary history among the observed individuals. The tips of a phylogenetic tree correspond to observed individuals and internal nodes represent common ancestors. We use the Newick notation to represent phylogenetic trees. For example, $(A, (B, C))$ represents the phylogenetic tree shown in Figure 1a.

As mentioned before, there is no perfect match between phylogenetic and transmission trees; for a given transmission tree, several phylogenetic trees are possible due to the possibility of different time orderings of infections.

**Example 1** *Because of different possibilities for orders of infections, two phylogenetic trees are consistent with the transmission tree $\{C \to A, C \to B\}$ (Figure 1d). The phylogenetic tree $(A, (B, C))$ (Figure 1a) corresponds to the case where $C$ first infects $A$ then $B$ i.e., $I_A > I_B$ while $(B, (A, C))$ (Figure 1f) corresponds to the case where $I_A < I_B$.*

In addition, $2^{n-1}$ transmission trees are possible for a given phylogentic tree with $n - 1$ internal nodes, one transmission tree for each possible labeling of internal nodes Kenah et al. (2016).

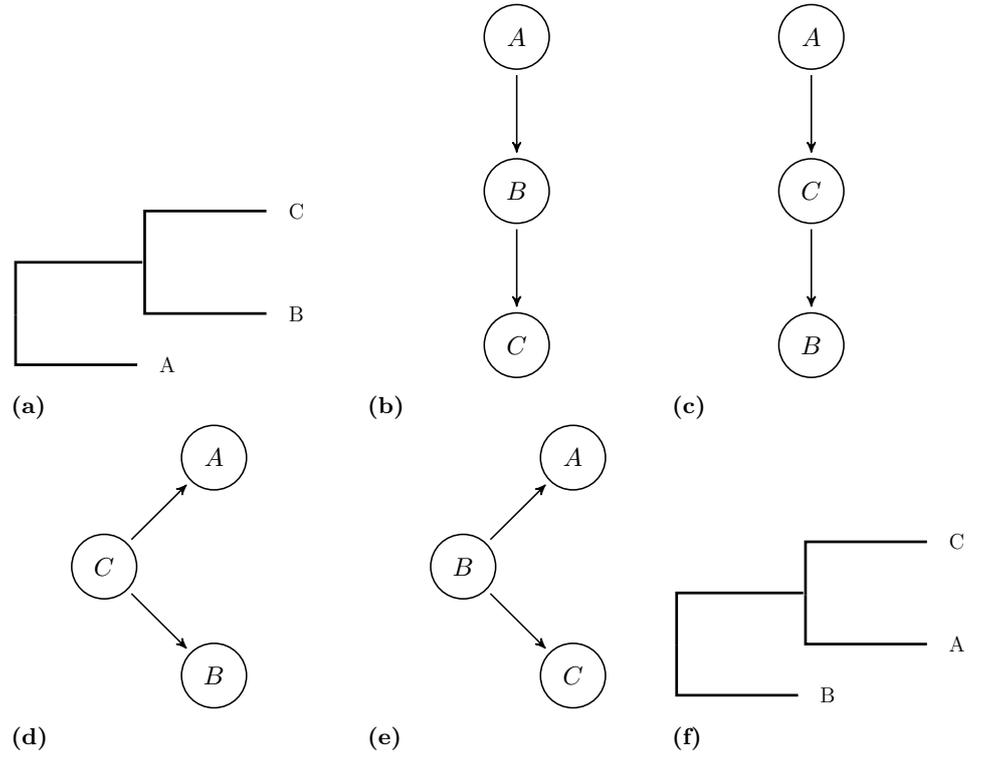

**Fig 1.** The graphical representations of a) the phylogenetic tree $(A, (B, C))$, transmission trees b) $\{A \to B, B \to C\}$, c) $\{A \to C, C \to B\}$, d) $\{C \to A, C \to B\}$, e) $\{B \to A, B \to C\}$ and f) the phylogenetic tree $(B, (A, C))$.

**Example 2** *For the phylogenetic tree $(A, (B, C))$ (Figure 1a), four transmission trees $\{A \to B, B \to C\}$ (Figure 1b), $\{A \to C, C \to B\}$ (Figure 1c), $\{C \to A, C \to B\}$ (Figure 1d), and $\{B \to A, B \to C\}$ (Figure 1e) are consistent with the given phylogenetic tree.*

However, provided infection times are known, a unique timed phylogenetic tree corresponds to a given transmission tree. Algorithms 1 and 2 provide a constructive two-step method to build the corresponding phylogenetic tree. In the first step, algorithm 1 reconstructs the topology of the phylogenetic tree (with equal branch lengths) and in the second step algorithm 2 assigns time points to all nodes of the reconstructed topology from the first step.

In algorithm 1, we start with a transmission tree with root $r$ and $k$ children namely $c_1, \ldots, c_k$. We assume the children's indexes are sorted by their infection times $I_{c_i} > I_{c_j}$ for $i < j$. In this recursive algorithm, we first construct a ladder-like mini-phylogenetic tree with $k + 1$ tips (similar to Figure 2). Let us assume that the tips of the mini-phylogenetic tree are indexed by their distances to the root such that the first tip is the closest one to the root and $(k+1)^{th}$ has the maximum distance in terms of the number of edges. In this case, we assign the transmission tree's root $(r)$ to the $(k+1)^{th}$ tip of the mini-phylogenetic tree. Then, we recursively construct a phylogenetic tree for each subtree $c_i$ of the transmission tree and place it at $i^{th}$ tip of the mini-phylogenetic tree. We illustrate in Figure 2b-d how this recursive algorithm works for the transmission tree shown in Figure 2a assuming $I_D \leq I_C \leq I_B$ and $I_F \leq I_E$. Once the phylogenetic topology is built, we need to assign branch lengths to the topology. For a leaf node, $t_x$ is equal to the sequencing time of $x$. For an internal node, we have $t_x = \min(m_l, m_r)$ where $m_l$ and $m_r$ are maximums of infection times for left and right subtrees of $x$, respectively. Algorithm 2 assigns time points to the nodes of the phylogenetic tree using an efficient postorder traversal. Then, the branch length of node $x$ is equal to $t_{\text{parent}(x)} - t_x$.

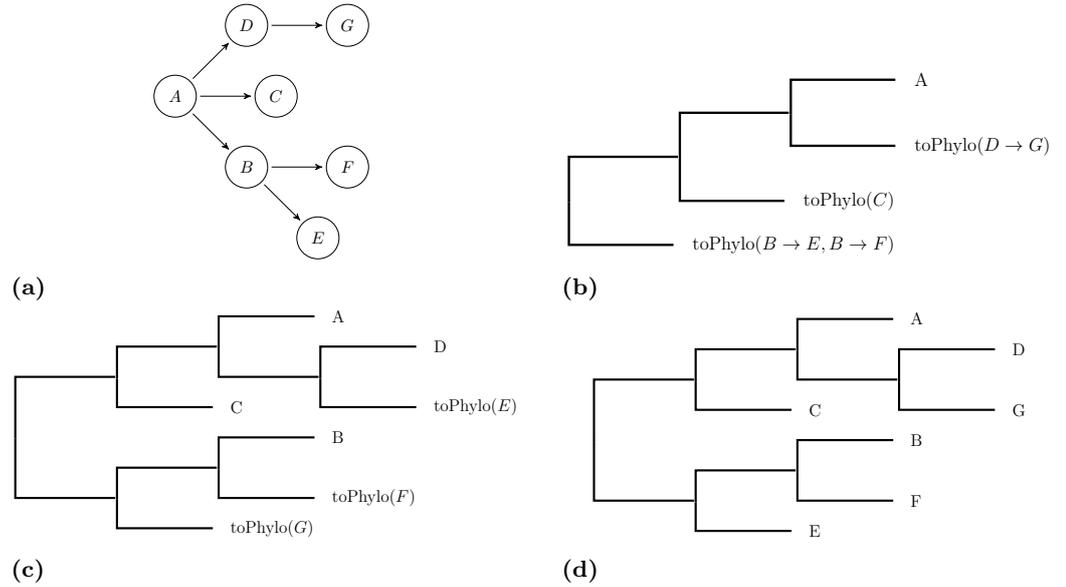

**Fig 2.** This figure illustrates several steps of the recursive algorithm 1 in reconstructing the phylogenetic tree corresponding to the transmission tree shown at part a. *toPhylo* is a recursive function converts a given transmission tree to the corresponding phylogenetic tree. We assume $I_D \leq I_C \leq I_B$ and $I_F \leq I_E$.

Now we propose a Bayesian method using Markov chain Monte Carlo (MCMC) for inferring transmission tree $(T)$, infection times $(I)$, and overall substitution parameter $(\alpha)$ given input data $D = \{g, l, r, s\}$ where $g = \{g_x \mid x \in \mathcal{P}\}$ denotes observed sequences, $s = \{s_x \mid x \in \mathcal{P}\}$ sequencing times, and $l = \{l_x \mid x \in \mathcal{P}\}$ and $r = \{r_x \mid x \in \mathcal{P}\}$ sets of left- and right-hand times of infection intervals, respectively. We incorporate infection intervals in estimation as soft constraints by adding a prior distribution on infection time $I_x$ such that it fulfills $P(l_x \leq I_x \leq r_x) = 0.95$. The posterior distribution is given by

$$P(T, I, \alpha \mid g, l, r, s) \propto P(g \mid T, I, s, \alpha) P(\alpha) P(T) P(I \mid l, r)$$

In order to compute $P(g \mid T, I, s, \alpha)$, we first identify the corresponding phylogenetic tree of transmission tree $T$ using algorithms 1 and 2 and then use the Felsenstein algorithm to compute the likelihood of the obtained phylogenetic tree Felsenstein (1981) based on Jukes and Cantor, 1969 (JC69) substitution model Jukes et al. (1969). The only parameter of the JC69 model is the overall substitution rate, $\alpha$. We use an improper uniform distribution for transmission tree, $P(T) = 1$ and an informative Gamma prior for $\alpha$ based on the available information on the substitution rate of the disease of interest. We assume $I \mid l, r \sim N(\frac{r+l}{2}, \sigma^2)$ where $\sigma = (l - r)/4$ (in order to fulfill the above-mentioned soft constraint). We use four moves to build an MCMC sampler to draw samples from the posterior distribution of transmission tree and parameters.

1. The first move is Subtree-Pruning-Regrafting (SPR) on the topology of the

**Algorithm 1** Reconstruction of phylogenetic tree topology for a given transmission tree and infection times.

---
**INPUT:** $tRoot$: the root of the transmission tree
**OUTPUT:** a phylogenetic tree with equal branch lengths

1: **function** TOPHYLO($tNode$)  ▷ a recursive function
2:     **if** $tNode.isLeaf$ **then**
3:         **return** new phyloNode(tNode.name)
4:     **end if**
5:     $cNodes \leftarrow$ sort($tNode.children$)  ▷ sort decreasingly by infection times
6:     pNode $\leftarrow$ new phyloNode()  ▷ create the root of mini-phylogenetic tree
7:     nextPNode $\leftarrow$ pNode
8:     **for** $i \leftarrow 1$ to $cNodes.size$ **do**
9:         nextPNode.right $\leftarrow$ toPhylo(cNodes[i])
10:         nextPNode.left $\leftarrow$ new phyloNode()
11:         nextPNode $\leftarrow$ nextPNode.left
12:     **end for**
13:     nextPNode $\leftarrow$ new phyloNode(tNode.name)
14:     **return** $pNode$
15: **end function**
16:
17: $toPhylo(tRoot)$  ▷ Function $toPhylo$ with the input $tRoot$ creates the corresponding phylogenetic tree.

---

**Algorithm 2** Assign time points to nodes of a phylogenetic tree topology given infection and sequencing times.

---
**INPUT:** phylogenetic tree topology P, infection times $I$ and sequencing times $S$
**OUTPUT:** assigned time points to the nodes of the input phylogenetic tree; denoted by $t_x$ for node $x$

1: **for** node x in postorder traversal of P **do**
2:     **if** $x$ is a leaf **then**
3:         $t_x \leftarrow S_x$
4:         $t_x^{\max} \leftarrow I_x$
5:     **else**
6:         $t_x \leftarrow \min\left(t_{x.\text{left}}^{\max}, t_{x.\text{right}}^{\max}\right)$
7:         $t_x^{\max} \leftarrow \max\left(t_{x.\text{left}}^{\max}, t_{x.\text{right}}^{\max}\right)$
8:     **end if**
9: **end for**

---

transmission tree. In this move, a subtree is selected and pruned from the transmission tree and then attached to a random node in the remaining tree. The default probability for choosing this move is 0.7.

2. The second move is a child-parent exchange. In this move, we exchange a random non-root node to its parent. The default probability for choosing this is 0.2.

3. The third move picks a random node of the transmission tree and updates its infection time using a uniform distribution on an interval consistent with infection times of other nodes in the tree. In particular, the new infection time should be smaller than its parent's infection time and larger than all its descendants'. The default probability for choosing this move is 0.1.

4. The last move updates the substitution parameter $\alpha$ using a lognormal random walk. This move is independent of other moves and is performed on every iteration.

## Results/Discussion

### Simulation study

This section assesses the performance of the proposed method in the reconstruction of transmission trees. For each tree size, we simulate a transmission tree using the susceptible-infected-removed (SIR) epidemic model. We choose the default values for the epidemic parameters according to a typical HIV outbreak: basic reproductive number (R0) 4, sequence length 3000, overall substitution rate ($\alpha$) $3 \times 10^{-3}$, according to the JC69 model. We choose the infection interval size as $W \times \mathrm{IQR}(I)$ where IQR(I) denotes inter-quartile region of infection times with the default value of 0.01 for $W$. Smaller value of $W$ provides more information about infection times.

We study four different sizes of infection intervals as $W = 0.01, 0.2, 0.5, 1000$. The infection interval for $W = 0.01$ is small i.e. infection times are almost exactly known. In contrast, there is almost no information available about infection times for extremely large $W = 1000$. In addition, we investigate different transmission trees with 15, 30, 50 and 100 nodes ($N$). For each parameter setting, we run the proposed MCMC sampler for 200000 steps. In order to estimate the posterior distribution of transmission trees, we choose 100 approximate i.i.d. samples from each chain after discarding the first quarter as burn-in. We build a consensus transmission tree from the obtained samples by the maximum parent credibility (MPC) algorithm Hall et al. (2015) and compare it to the underlying true transmission tree by reporting the number of shared edges between the two trees. For each parameter setting, we repeat the outbreak simulation and the MCMC sampler for 20 times (Figure 3). In addition, we compare the performance of the proposed method to *phybreak*, a recent computational method for reconstruction of transmission trees Klinkenberg et al. (2016). This method was shown to outperform two other computational tools implemented in R packages *Outbreaker* Jombart et al. (2014) and *TransPhylo* Didelot et al. (2014); hence we only compare the proposed method to *phybreak*. Figure 3 displays that infection times coupled with genetic sequence are necessary for accurate reconstruction of transmission trees. In particular, for different values of $N$, the proposed method is able to recover 90% of true transmission trees for $W = 0.01$; by contrast, the performance of the method is highly unreliable for $W = 1000$ and comparable to the performance of *phybreak* (Figure 3). In addition, We perform a sensitivity analysis, which demonstrates that the presented method has a robust performance for a broad range of parameter values (supplementary Figure S1).

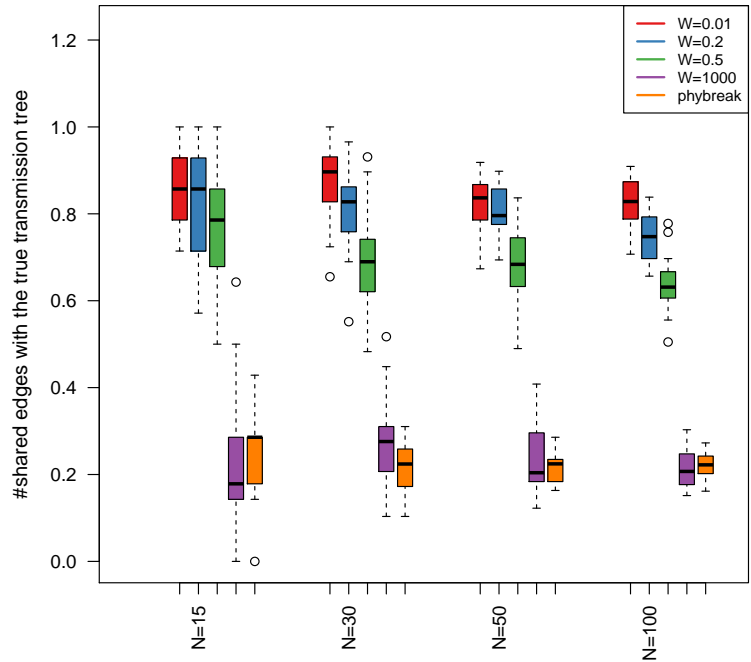

**Fig 3.** Performance of the proposed MCMC sampler for different sizes of transmission trees ($N$) and various infection time intervals (specified by $W$). Performance is also compared to an alternative reconstruction method, *phybreak*, which does not take into account infection intervals. The proposed method is able to incorporate this additional information in the estimation. According to this figure, availability of more accurate data on infection times leads to more accurate reconstruction of transmission trees. Both the proposed method with large $W$ and the *phybreak* method provide highly unreliable estimates.

## HIV Application

This section applies the proposed method in reconstructing a transmission tree for an HIV dataset from San Diego, California, which contains information on 19 subjects whose sequences are found in the largest transmission cluster reported inLittle et al. (2014). An estimated date of infection (EDI) is available for each sequence, based on the methods described in Le et al. (2013). To account for uncertainty of EDI, we assume an infection time occurred within a six-month interval centered at the corresponding EDI. We run the proposed MCMC for 200,000 iterations and obtain 50 approximately i.i.d. transmission trees from the MCMC chain after discarding the first one third as the burn-in phase. Two similar transmission trees that appeared in the thinned MCMC chain are shown in Figure 4. Using a one-year interval for infection times results in 41 unique transmission trees in the thinned MCMC chain. Each node of the transmission trees represents an HIV patient either belongs to low viral load (VL) category, defined as less than $10^5$cps/ml, or with higher VL ($\geq 10^5$cps/ml).

To test for dependence of the probability of linkage on whether the nodes are in the same VL category, we use a multiple imputation (MI) framework, in which the transmission trees are treated as the missing data. The data consists of $(W_{obs}, W_{mis})$ where $W_{obs}$ denotes VL category and $W_{mis}$ denotes the transmission tree. We obtain $M$ draws (imputations) from the posterior predictive distribution of $W_{mis}$ from the MCMC chain. To test the null hypothesis that linkage is independent of VL category, we define Q a test statistic as $Q = O - E$ where O is the number of links between low VL nodes and E is the expected number of such links under the null. $\hat{Q}^{(m)}$ denotes the computed quantity of interest for m$^{th}$ imputed dataset; we define $\tilde{Q}$ as the sample mean of $\hat{Q}^{(m)}$ across $M$ imputations. To test the null hypotheses we consider two approaches: One is to calculate exact p-values for each tree and then marginalize across the trees by taking the sample average across them Wang et al. (2010). The second is to calculate the variance of $\tilde{Q}$ is a combination of within- and between-imputation variances Kenward and Carpenter (2007); Lynch and DeGruttola (2015), $\text{var}(\tilde{Q}) = (1 + M^{-1})B + U$ where

$$B = \frac{1}{M-1} \sum_{m=1}^{M} \left( \hat{Q}^{(m)} - \tilde{Q} \right)^2 \quad (1)$$

and

$$U = \frac{1}{M} \sum_{m=1}^{M} \text{var}\left( \hat{Q}^{(m)} \right) \quad (2)$$

To approximate the variance of var($\hat{Q}$) under the null hypothesis, we obtain empirical variance conditional on each tree obtained through permutation. The test statistic has an asymptotic $t$ distribution for which the degrees of freedom are calculated as explained in Kenward and Carpenter (2007). The first method of averaging p-values across draws of transmission trees (imputations) yields a p-value of 0.77. The p-value associated with the second, asymptotic, method is 0.75. In addition to viral load, we also looked at the effect of individual characteristics on probability of linkage to others who share them; these characteristics included Hispanic ethnicity (4 of 19 subjects) and number of sexual partners (dichotomized as 1, 2, or 3, and more than 3). The p-values for the two methods were 0.14 and 0.11 for Hispanic ethnicity and 0.35 and 0.28 for number of partners.

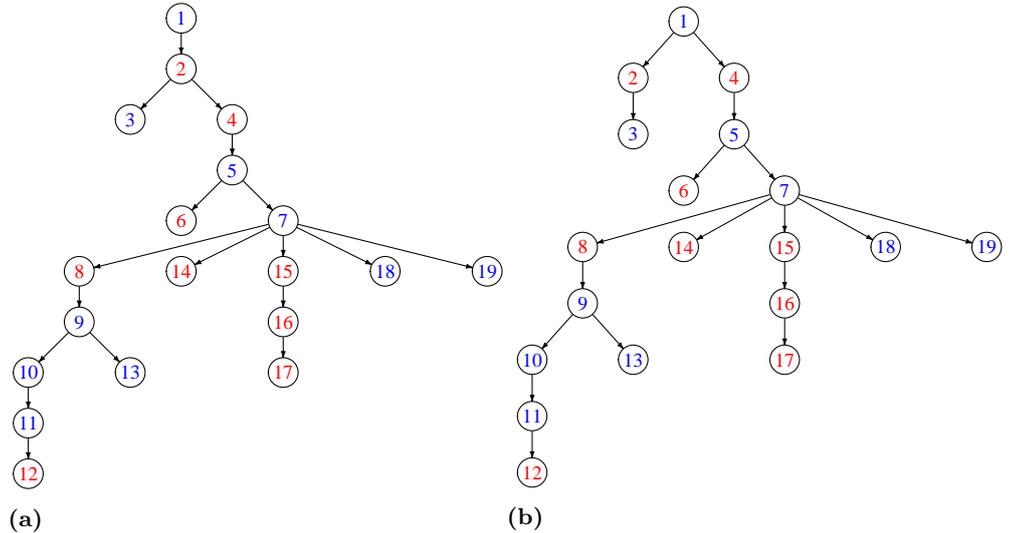

**Fig 4.** Two similar transmission trees appeared in the thinned MCMC chain using a six-month interval for an HIV dataset from San Diego with relative frequencies a) 88% and b) 12%. Blue nodes represent HIV patients with low viral load (VL) defined as less than $10^5$cps/ml and red nodes show those with higher VL ($\geq 10^5$cps/ml).

**Table 1.** This table provides information about VL level and the number of transmissions of each patient. The number of transmissions is estimated by outdegree of each node (patient) averaged over sampled transmission trees from the MCMC chain.

| Patient | VL(High/Low) | VL(cps/ml) | Average degree |
|---------|--------------|------------|----------------|
| 1  | High | 112000  | 0.00 |
| 2  | High | 179000  | 1.00 |
| 3  | High | 431000  | 1.00 |
| 4  | High | 504000  | 1.00 |
| 5  | High | 552000  | 1.88 |
| 6  | High | 750000  | 0.00 |
| 7  | High | 750000  | 0.00 |
| 8  | High | 1764120 | 0.00 |
| 9  | High | 3770000 | 0.00 |
| 10 | High | 7300000 | 1.00 |
| 11 | Low  | 1910    | 2.00 |
| 12 | Low  | 6265    | 1.12 |
| 13 | Low  | 14700   | 5.00 |
| 14 | Low  | 21200   | 0.00 |
| 15 | Low  | 26300   | 2.00 |
| 16 | Low  | 30200   | 0.00 |
| 17 | Low  | 45300   | 1.00 |
| 18 | Low  | 59500   | 1.00 |
| 19 | Low  | 73200   | 0.00 |

In addition, the relationship between outbound edges and VL level is shown in Table 1. The number of outbound edges is associated with the ability to transmit HIV. Average degree of each node is computed using 50 transmission trees obtained from the MCMC chain. We use the same statistical tests to compare the average number of outbound edges in each VL group. Testing the null hypothesis of no difference between outdegrees of low VL patients versus high VL patients by averaging p-values across imputations yields a p-value of 0.11. The asymptotic approach described above yields a p-value=0.09. These analyses do not provide sufficient evidence against the null hypothesis to conclude that outdegrees of low VL and high VL patients are different, but suggest that this question may warrant further study. Using the same approach, we found not effect of Hispanic ethnicity or number of partners on outbound edges.

## Conclusion

Transmission trees provide more detailed information about spread of epidemic diseases than phylogenies. However, accurate reconstruction of transmission chains using genetic sequence data is challenging. This paper investigated inference issues and proposed a new method for investigating features of the trees; our results suggest that sequence data must be augmented by information regarding infection times for reliable reconstructions of underlying transmission trees. We have introduced a novel Bayesian inference method for reconstruction of transmission trees using these augmented data. Simulation studies showed that the accuracy of the presented method improves as the uncertainty on infection times decreases. One limitation of the proposed method is that it works under the assumption that sequence data is available for all patients. This assumption is reasonable for outbreaks in closed communities such as in prison or hospital. However, further research is required to develop a Bayesian inference method when only sequences for a subset of patients are available. Another promising area of further research is to expand this framework to take into account within-host diversity in patients, which requires next-generation sequencing data and raises additional questions such as the subset of transmitted viral variants.

## Acknowledgement

We would like to thank Dr. Davey Smith for his contribution in providing the HIV data and his insightful comments (grant number: CFAR AI03621).

## Funding

This work was supported by the National Institutes of Health under award numbers AI106039, GM110749 (to Susan Little) and R37 AI 51164 (to Hesam Montazeri and Victor DeGruttola). The funders had no role in study design, data collection and analysis, decision to publish, or preparation of the manuscript.

# Supporting information

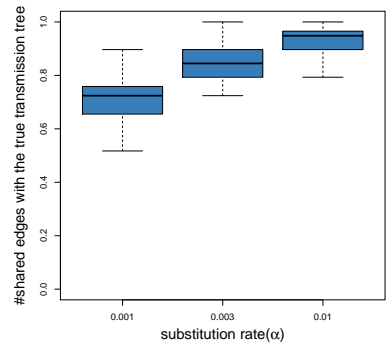

(a)

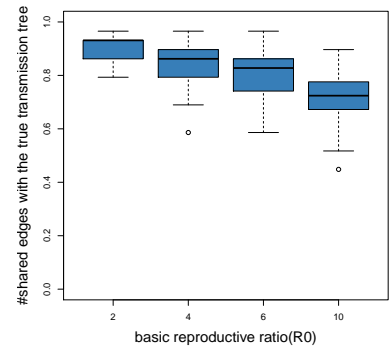

(b)

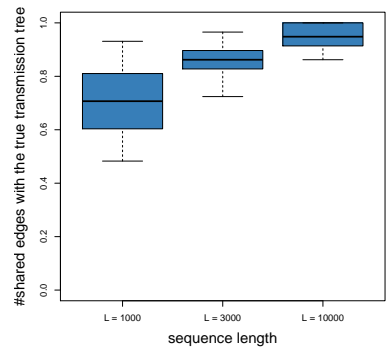

(c)

**Fig S1.** Performance of the proposed method in transmission tree reconstruction against different parameters a) substitution rate b) basic reproductive number c) sequence length.